\documentstyle[aps,twocolumn]{revtex}

\begin{document}

\input epsf
\newcommand{\infig}[2]{\begin{center}\mbox{ \epsfxsize #1
                       \epsfbox{#2}}\end{center}}

\newcommand{\be}{\begin{equation}}
\newcommand{\ee}{\end{equation}}
\newcommand{\bea}{\begin{eqnarray}}
\newcommand{\eea}{\end{eqnarray}}
\newcommand{\wee}[2]{\mbox{$\frac{#1}{#2}$}}   
\newcommand{\unit}[1]{\,\mbox{#1}}
\newcommand{\degree}{\mbox{$^{\circ}$}}
\newcommand{\ltish}{\raisebox{-0.4ex}{$\,\stackrel{<}{\scriptstyle\sim}$}}
\newcommand{\vs}{{\em vs\/}}
\newcommand{\bin}[2]{\left(\begin{array}{c} #1 \\ #2\end{array}\right)}
                        
\draft

\title{Creation of coherence in Bose-Einstein condensates by atom detection}

\author{Peter Horak and Stephen M.\ Barnett}
\address{Department of Physics and Applied Physics,
University of Strathclyde, Glasgow G4 ONG, United Kingdom}
\date{\today, submitted to Phys.\ Rev.\ A}

\maketitle

\begin{abstract}
We investigate the creation of a relative phase between two Bose-Einstein
condensates, initially in number states, by detection of atoms and show how
the system approaches a coherent state. Two very distinct time scales are
found: one for the creation of the interference is of the order of the
detection time for a few single atoms and another, for the preparation of
coherent states, of the order of the detection time for a significant
fraction of the total number of atoms. Approximate analytic solutions are
derived and compared with exact numerical results.
\end{abstract}


\narrowtext


\section{Introduction}

The first experimental realisations of Bose-Einstein condensation in
dilute atomic gases \cite{bec95a,bec95b,bec95c}
have opened up a new field in atomic and quantum physics. Despite its apparent
complexity, the condensate is well-described by a macroscopic wavefunction, or
complex field, obeying a nonlinear wave equation (the Gross-Pitaevskii
equation). The nature of this complex field and in particular its apparent
phase has been the subject of some debate.

One specific topic which has been frequently investigated is the question of
the coherence properties \cite{Burt,Ketterle}
 of Bose-Einstein condensates as established in
interference experiments \cite{Andrews,Hall}. 
The underlying question is whether the phase or, more
precisely, the relative phase of two condensates is created by a spontaneously
broken symmetry \cite{Leggett} or by some other mechanisms \cite{Kagan}. 
It has been
suggested \cite{Javanainen} (and studied in detail by several authors 
\cite{Wong,Cirac,Castin,Graham,Ruostekoski,Sinatra})
that such a relative phase can be created by the
detections of individual atoms in an interferometric setup, that is, where the
origin of the detected atoms is intrinsically unknown. This process gives rise
to a definite (but unpredictible) relative phase of two condensates even if the
initial states of the condensates are of undefined phase, as for initial number
states \cite{Javanainen,Wong,Castin}, initial Poissonian states
\cite{Cirac,Graham}, or initial thermal states \cite{Graham},
by entangling the states of the two condensates.

In this work we will study in detail the creation of coherence between
two condensates which initially have well defined occupation numbers, not only
in the limit where the number of atoms detected is small compared to the total
number of atoms but also in the long time limit. We concentrate thereby on two
manifestations of coherence: the creation of interference fringes and
the evolution of the compound two-condensate system towards a kind of
coherent state. We start with an idealized model 
(Secs. \ref{sec:interference}-\ref{sec:state}) and later introduce more
realistic features including atomic collisions 
(Sec.~\ref{sec:general}). 
We use exact numerical simulations combined with
approximate analytical solutions to determine
the time evolution of our system. We show
that these two measures of coherence are created on very different time scales.
The relative phase associated with the appearance of interference fringes
develops in a time of order $1/(N\gamma)$, where $N$ is the
total initial number of atoms in the two condensates and $\gamma$ is the
rate at which atoms leak out of the condensates and are detected.
The approach towards coherent states, in contrast, takes a larger time of order
$1/\gamma$. These two timescales correspond to timescales recently identified
for the interaction of a Bose-Einstein condensate or other bosonic system with
its environment \cite{Barnett}. In this case, a well-specified atom number will
change on the timescale $1/(N\gamma)$ but any coherence present decays on a
timescale $1/\gamma$.


\section{Model}

\begin{figure}[tb]
\infig{18em}{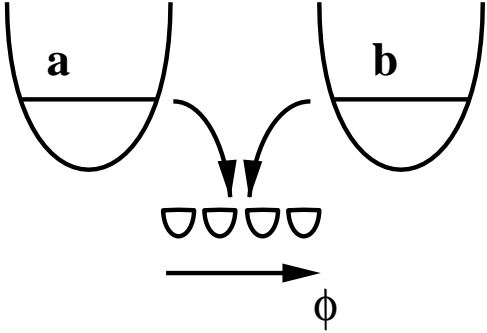}
\caption{Schematic representation of the interfering Bose-Einstein condensates.}
\label{fig:scheme}
\end{figure}

Let us first introduce the model system which we use to investigate the
coherence properties in Bose-Einstein condensation
\cite{Javanainen,Wong,Cirac,Castin,Graham,Ruostekoski,Sinatra}.
We consider two independent non-interacting single-mode Bose-Einstein
condensates with creation (annihilation) operators $a^\dagger$ ($a$) and 
$b^\dagger$ ($b$), respectively. Atoms are leaking out of the condensates
and are detected individually and spatially resolved. The same
detectors simultaneously monitor decays from both condensates, that is, atoms
coming from different condensates are allowed to interfere, see 
Fig.~\ref{fig:scheme}.
Thus, the detections are described by the annihilation operators
\be
a(\phi) = \frac{1}{\sqrt{2}} \left( a + b e^{-i\phi} \right)
\ee
where $\phi \in [-\pi,\pi]$ is related to the position of the detector $x$ by 
$\phi=px/\hbar$ where $p$ is the momentum of the atoms leaking out of the
condensates.

Assume now that the system can be described at a certain time $t$ by a wave
function $|\psi\rangle$. Then the probability of detecting the next atom at
position $\phi$ is given by
\be
P(\phi) = {\cal N} \langle \psi|a(\phi)^\dagger a(\phi)|\psi \rangle
\ee
where the normalisation constant ${\cal N}$ is chosen such that 
\be
\int_{-\pi}^{\pi} d \phi \, P(\phi) = 1.
\ee
This probability function can be rewritten as
\be
P(\phi) = \frac{1}{2\pi}\left[ 1+\beta_c \cos(\phi-\theta) \right]
\label{eq:prob}
\ee
where
\be
\beta_c = \frac{2|\langle \psi|a^\dagger b|\psi \rangle|}
{\langle \psi|a^\dagger a + b^\dagger b|\psi \rangle}
\label{eq:beta}
\ee
is the visibility of the interference fringes conditioned 
on the quantum state $|\psi\rangle$, and $\theta$ gives the most likely
position of detection of the next atom. It should be emphasized that this
visibility is not the one obtained by detecting a large number of atoms from the
initial state $|\psi\rangle$, but the one obtained by preparing this state
$|\psi\rangle$ very often and measuring a {\em single\/} atom in each run.
This difference is important since every detection changes the state of the
system and thus changes the conditional visibility $\beta_c$.

This system has already been investigated by several authors 
\cite{Javanainen,Wong,Cirac,Castin,Graham,Ruostekoski,Sinatra}
in order to show that the
detection of atoms breaks the underlying symmetry of the system and thus creates
interference fringes. This is true even if the {\em initial\/} 
state of the system does not exhibit any prefered phase so that $\beta_c=0$.
It is the main purpose of our
work to quantify this creation of coherence between two initially
uncorrelated Bose-Einstein condensates.


\section{Creating interference from initial number states}
\label{sec:interference}

In this section we will discuss the creation of interference fringes as a
consequence of consecutive detections of atoms when the two condensates are
initially in number states. The full system is given initially by
the quantum state
\be
|\psi_0\rangle = |n_1,n_2\rangle.
\ee
After the detection of $k$ atoms at positions $\phi_1$, $\phi_2$, ..., 
$\phi_k$ the (unnormalized) state of the system is then
\be
|\psi_k\rangle = (a+be^{i\phi_k})\dots (a+be^{i\phi_1})|\psi_0\rangle.
\ee
The conditional probability of detecting the $(k+1)$th atom at 
$\phi_{k+1}$ then reads
\bea
& & P(\phi_{k+1}|\phi_k,...,\phi_1) = \nonumber \\
& & \quad\quad = \frac{1}{2\pi}
  \frac{\langle\psi_k|(a^\dagger+b^\dagger e^{-i\phi_{k+1}})
     (a+b e^{i\phi_{k+1}}) |\psi_k\rangle}
     {\langle\psi_k|a^\dagger a + b^\dagger b |\psi_k\rangle} 
     \nonumber \\
& & \quad\quad =  \frac{1}{2\pi} \frac{\langle\psi_{k+1}|\psi_{k+1}\rangle}
     {(N-k)\langle\psi_k|\psi_k\rangle}
\eea
where $N=n_1+n_2$. Thus, the probability for the sequence of detections 
$\phi_1$, $\phi_2$, ..., $\phi_k$ is
\bea
& & P(\phi_k,...,\phi_1) = P(\phi_1)P(\phi_2|\phi_1)... = \nonumber \\
& & \quad\quad =
  \frac{1}{(2\pi)^k} \frac{\langle\psi_k|\psi_k\rangle}{N(N-1)...(N-k+1)}.
\eea
The conditional visibility $\beta$ (we will write $\beta$ instead of $\beta_c$
in this section in order to simplify the notation)
for the state after $k$ detections is
\bea
\beta = \frac{2|\langle\psi_k|a^\dagger b |\psi_k\rangle|}
     {\langle\psi_k|a^\dagger a + b^\dagger b |\psi_k\rangle} 
     = \frac{2|\langle\psi_k|a^\dagger b |\psi_k\rangle|}
     {(N-k)\langle\psi_k|\psi_k\rangle}.
\label{eq:betak}
\eea
Thus the average visibility after $k$ detections is
\be
\langle \beta\rangle_k = \int d\phi_1...d\phi_k \frac{2}{(2\pi)^k}
   \frac{(N-k-1)!}{N!} |\langle\psi_k|a^\dagger b |\psi_k\rangle|.
\label{eq:bmean}
\ee
Note, however, that this mean conditional visibility is rather difficult 
to access experimentally since it involves the averaging over 
{\em ensembles\/} of experiments where each ensemble consists of repeatedly
preparing the quantum state of the system by the {\em same\/} sequence of
detections from the same initial number state and measuring the position
of the next detection. Nevertheless, this proves to be a useful measure
theoretically especially to describe the time scale on which interference
is created as we will show later in this section.

For the first two detections, that is, $k=1$, $2$, the integral in 
Eq.~(\ref{eq:bmean}) can be evaluated analytically with the results
\bea
\langle \beta\rangle_1 &=& \frac{2n_1 n_2}{(n_1+n_2)^2-(n_1+n_2)}, 
   \label{eq:beta1} \\
\langle \beta\rangle_2 &=& \frac{4}{\pi} \langle \beta\rangle_1.
   \label{eq:beta2}
\eea
For $n_1=n_2=N/2$ we thus obtain \cite{Wong,Graham}
\be
\langle \beta\rangle_1 = \frac{1}{2}\frac{1}{1-1/N}.
\ee
For any initial number of atoms this is larger than $1/2$, meaning that the
first detection already increases the average visibility from zero to more than
half its maximum value of one.

For $k>2$ Eq.~(\ref{eq:bmean}) can no longer be evaluated analytically but we
will derive an approximate solution in the following. 
Let us assume that the system starts in the quantum state 
$|\psi_0\rangle = |n,n\rangle$ (same number of atoms in both condensates) and
all detections occur at the same position $\phi$. Of course, this is a highly
unlikely detection sequence, but this assumption gives surprisingly good 
approximate results as we will see. Without loss of generality we may assume
$\phi=0$. Thus
\be
|\psi_k\rangle = (a+b)^k |\psi_0\rangle.
\label{eq:psikeq}
\ee
From this we obtain the norm of the state as
\bea
\langle \psi_k |\psi_k\rangle &=& \sum_{m=0}^k \bin{k}{m}^2
   \langle n,n|(a^\dagger)^m a^m (b^\dagger)^{k-m} b^{k-m}|n,n\rangle 
      \nonumber \\
   &=& \frac{n!^2}{(2n-k)!} \sum_{m=0}^k \bin{k}{m}^2 \bin{2n-k}{n-m}.
\eea
We now approximate the binomial coefficients by Gaussians using
\be
\bin{k}{m}\approx 2^k\sqrt{\frac{2}{\pi k}}e^{-\frac{2}{k}(m-k/2)^2}
\label{eq:binapprox}
\ee
and replace the sum over $m$ by an integral from $-\infty$ to $\infty$
which yields
\be
\langle \psi_k |\psi_k\rangle \approx \frac{n!^2}{(2n-k)!} 2^{2n+k}
   \frac{2}{\pi} \frac{1}{\sqrt{k(4n-k)}}.
\label{eq:psinorm}
\ee
Analogously we obtain
\be
\langle \psi_k |a^\dagger b|\psi_k\rangle \approx \frac{n!^2}{(2n-k-1)!} 
   2^{2n+k-1}\frac{2}{\pi} \frac{e^{-1/k}}{\sqrt{k(4n-k-2)}} 
\ee
and thus from Eq.~(\ref{eq:beta}) the final result
\be
\beta_k \approx e^{-1/k}
\label{eq:bapprox}
\ee
independent of $n$ (to order $1/n$). Hence, independent of the initial number of
atoms in the condensates, it always needs the same (and very small) number of
detected atoms to create the interference. If the initial total number
of atoms is $N$ and each atom decays with a rate $\gamma$ out of the
condensate, then the time to create the interference pattern will thus be of the
order of $1/(N\gamma)$.

\begin{figure}[tb]
\infig{20em}{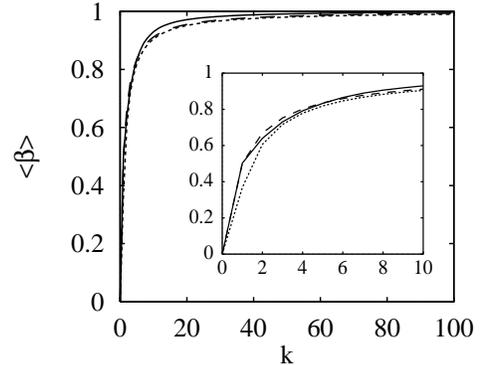}
\caption{Average conditional visibility $\langle \beta\rangle$ vs number of
detected atoms $k$: exact numerical solution (solid
curve), numerical solution if all atoms are detected at the same position
(dashed curve), approximate analytical solution (dotted curve). The initial
state is $n_1=n_2=100$.}
\label{fig:beta1}
\end{figure}

In Fig.~\ref{fig:beta1} we compare the approximate result of 
Eq.~(\ref{eq:bapprox}) with the exact
numerical solution of Eq.~(\ref{eq:betak}). The comparison shows that the
approximation is excellent for $k$ as low as 2 (10\% difference), 
3 (5\%) and 4 (3\%) for $n=100$, even if the approximation made in
Eq.~(\ref{eq:binapprox}) only holds for $k\gg 1$. 
We also plot the numerical solution of 
Eq.~(\ref{eq:bmean}), that is, the visibility averaged over all possible outcomes
with the appropriate probabilities which we obtained by Monte-Carlo simulations
of the detection process. We see that the difference in the creation of
interference fringes between the case where all atoms are detected at the same
place and the case where all possible detection positions are allowed is
relatively small. 
This is somewhat surprising given the great difference between the extremely
peaked distribution of detection positions of our approximation compared with
the expected sinusoidal behavior [see Eq.~(\ref{eq:prob})]. It is, however, a
consequence of the fact that only a small number of detections determine the
interference pattern for all subsequent detections.

So far we have restricted ourselves to the case of equal initial occupation
numbers of the two condensates. In the case of unequal initial numbers the
approximation assuming that all the detections occur at the same position
fails as it turns out that this changes the relative occupation of the two
condensates (which is constant if the condensate decay rates are equal).
An analytic approximation can be found, however, assuming that the number of
detected atoms $k$ is small compared to the total number of atoms $N$. 
In this case the normalisation of the state after $k$ equal detections is
\bea
& & \langle \psi_k |\psi_k\rangle = \nonumber \\
& & \quad =\sum_{m=0}^k \bin{k}{m}^2
   \langle n_1,n_2|(a^\dagger)^m a^m (b^\dagger)^{k-m} b^{k-m}|n_1,n_2\rangle
      \nonumber \\
& & \quad \approx \sum_{m=0}^k \bin{k}{m}^2 n_1^m n_2^{k-m}.
\eea
The latter expression can be approximated by replacing the binomials with
Gaussians and the sum by an integral as before. Applying the same procedure
to $\langle \psi_k |a^\dagger b|\psi_k\rangle$ one finally obtains the
conditional visibility
\be
\beta_k \approx \frac{2\sqrt{n_1 n_2}}{n_1+n_2} e^{-1/k}
\label{eq:buneq}
\ee
the long-time limit of which ($k \gg 1$) has also been found in
Ref.~\cite{Graham}.

Three features are worth mentioning here. First, the evolution of the visibility
as a function of the number of detected atoms is the same as in the case of
equal initial atom numbers. Hence, also in this case only a few detections are
required to establish the interference pattern and hence the corresponding time
scale is again given by $1/(N\gamma)$. Second, the maximum possible visibility
is reduced to a value which depends on the ratio of the initial occupation
numbers of the two condensates. Third, this maximum visibility is exactly
the same as for initial {coherent\/} states of the condensates with the same
mean numbers of atoms.

\begin{figure}[tb]
\infig{20em}{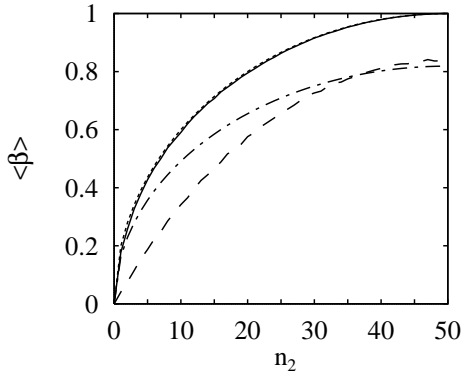}
\caption{Average conditional visibility $\langle \beta\rangle$ vs number $n_2$
of atoms initially in one of the condensates, the total number is 
$n_1+n_2=100$. Solid curve: exact numerical result after 99 detections, dashed:
exact result after 5 detections, dotted: approximate analytic result after 99
detections, dash-dotted: approximate result after 5 detections.}
\label{fig:beta2}
\end{figure}

In Fig.~\ref{fig:beta2} we compare these results with the exact numerical
solutions for arbitrary detection positions. We see that in the
long-time limit, with nearly all of the atoms detected, the agreement is exact,
thus the visibility approaches the value found for coherent states of the same
mean atom number. After a small number $k$ of detections a more significant
deviation from our approximate result is found, especially for significantly
differing
initial occupation numbers $n_1$ and $n_2$. However, we still find that
Eq.~(\ref{eq:buneq}) predicts the correct order of magnitude for the time
required to establish the interference.


\section{Preparation of coherent states from initial number states}
\label{sec:state}

In this section we will discuss how the system state approaches a coherent state
in course of the sequence of detections. However, in order to simplify the
discussion and the numerical simulations we will assume that the {total\/}
number of atoms in the two condensates is exactly known at any time; the
system is initially in a number state $|n_1,n_2\rangle$ with $n_1$ atoms in
condensate $a$ and $n_2$ atoms in condensate $b$ and the number $k$ of detected
atoms is known at any time. Thus the system will never approach a 
harmonic oscillator coherent state, that is, an eigenstate to the
annihilation operators $a$ and $b$, since these are
superposition states of different total atom numbers. Instead it will be
shown that the system approaches a state which can be described as the
restriction of a coherent state to the subset of states with a fixed total
number of atoms. Hereafter we will refer to these states as 
``atomic coherent states'' as they are a representation of the atomic coherent
states \cite{states}. We can define these states to be
\be
|\mu,\nu\rangle_N = \frac{1}{\sqrt{N!}} 
   \left( a^\dagger \mu + b^\dagger \nu\right)^N |0,0\rangle
\label{eq:qcs}
\ee
where $\mu$ and $\nu$ obey the relation $|\mu|^2+|\nu|^2=1$.
Some properties of these atomic coherent states and their
relation to the coherent states are discussed in Appendix~\ref{appendix}.

The specific case of these states with $\mu=e^{i\phi}/\sqrt{2}$ and
$\nu=e^{-i\phi}/\sqrt{2}$ has also been refered to as ``phase states'' 
\cite{Castin,Sinatra},
\be
|\phi\rangle_N = \frac{1}{\sqrt{2^N N!}} 
   \left( a^\dagger e^{i\phi} + b^\dagger e^{-i\phi}\right)^N |0,0\rangle
\label{eq:phs}
\ee
since their conditional visibility is $\beta_c=1$.

The problem which we will consider in this section is the following. Assume the
system is initially in the number state $|\psi_0\rangle=|n_1,n_2\rangle$.
Then $k$ atoms are detected at positions $\phi_1$, $\phi_2$, ..., $\phi_k$
and the resulting state $|\psi_k\rangle$ is analyzed. What is the probability
of finding an atomic coherent state $|\mu,\nu\rangle_{N-k}$, where
$N=n_1+n_2$? To answer
this we have to evaluate the probability function
\be
P(\mu,\nu)=\frac{|\langle \psi_k|\mu,\nu\rangle_{N-k}|^2}
             {\langle \psi_k|\psi_k\rangle},
\ee
or equivalently
\be
P(\phi)=\frac{|\langle \psi_k|\phi\rangle_{N-k}|^2}
             {\langle \psi_k|\psi_k\rangle}
\ee
in the case of $n_1=n_2=N/2$.

Let us consider first the case of equal initial atom numbers $n_1=n_2=n$ and
assume without loss of generality that the first atom is detected at position
$\phi_1=0$ so that $|\psi_1\rangle=(a+b)|n,n\rangle$. We then find
\be
P(\phi) = \frac{1}{2^{2n}} \bin{2n}{n} (1+\cos 2\phi)
        \approx \frac{1}{\sqrt{\pi n}} (1+\cos 2\phi)
\ee
where for the last approximation we have again used Eq.~(\ref{eq:binapprox}).
One can easily check by numerical simulation that this approximation is 
highly acurate even
for just a few atoms in the condensates. Hence the overlap of the state after
one detection with any phase state is very small of the order of $1/\sqrt{n}$.

For the analysis of the state after $k$ detections we will assume, as in the
previous section, that all atoms are detected in the same position, so that
$|\psi_k\rangle = (a+b)^k |\psi_0\rangle$, and will compare the results for
$P(\phi)$ with numerical simulations at the end.
The state overlap after $k$ detections is then
\bea
\langle \psi_k|\phi\rangle_{2n-k} &=& \frac{1}{\sqrt{2^{2n-k}(2n-k)!}}
      \nonumber \\
 & & \times   \langle n,n|(a^\dagger+b^\dagger)^k
    (a^\dagger e^{i\phi}+b^\dagger e^{-i\phi})^{2n-k}|0,0\rangle \nonumber \\
    & = & \frac{n!}{\sqrt{2^{2n-k}(2n-k)!}} \nonumber \\
 & & \times \sum_{p=0}^k \bin{k}{p} \bin{2n-k}{n-p} e^{i\phi(k-2p)}
\eea
which, after applying the approximation of Eq.~(\ref{eq:binapprox}), becomes
\be
\langle \psi_k|\phi\rangle_{2n-k} \approx \frac{1}{\sqrt{2^{2n-k}(2n-k)!}}
   \frac{n!2^{2n}}{\sqrt{\pi n}} 
   e^{-\frac{\phi^2}{4}\frac{k(2n-k)}{n}}.
\ee
This, together with the normalization factor, Eq.~(\ref{eq:psinorm}), yields
\be
P(\phi) \approx  \frac{1}{2}\sqrt{\frac{k(4n-k)}{n^2}}
   e^{-\frac{\phi^2}{2}\frac{k(2n-k)}{n}}.
\label{eq:papprox}
\ee
In the limit of $k\ll n$ this result is consistent with the results presented
in Refs.~\cite{Cirac,Castin,Ruostekoski}.

\begin{figure}[tb]
\infig{20em}{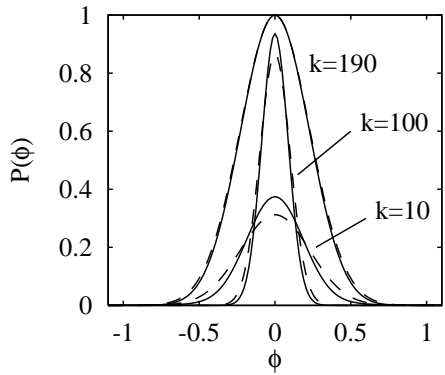}
\caption{Probability $P(\phi)$ of finding the phase state $|\phi\rangle$ 
after $k$ atom detections from initial state $|n_1=100,n_2=100\rangle$. The
solid curves correspond to exact numerical solutions 
(averaged over 1000 Monte-Carlo simulations) and the dashed curves to
the analytic approximation given in Eq.~(\protect\ref{eq:papprox}).}
\label{fig:pphi1}
\end{figure}

In Fig.~\ref{fig:pphi1} we compare this analytic approximation for the
probability function $P(\phi)$ with the exact numerical solution obtained by
Monte-Carlo simulations where all possible detection positions are taken into
account. (On the scale of Fig.~\ref{fig:pphi1} the curves for the approximate
solution and the exact solution if all atoms are detected at the same position
coincide almost exactly.) After only a few detections ($k=10$ in the figure) the
probability function is already well approximated by a broad Gaussian with a
relatively small maximum, so the overlap with any phase state is still small
at this time. However, we note that the maximum overlap is {\em larger\/} than
predicted by the approximate analytic solution, which was derived under the
assumption that all atoms are detected at the same position. This may seem
surprising but can be explained by the fact that such a highly peaked position
distribution of the detected atoms is far from the one expected for a coherent
state. After the detection of half of the atoms ($k=100$) the maximum overlap
with a phase state is already close to one and the width of the Gaussian has
decreased significantly. For a larger number of detected atoms ($k=190$) the
maximum overlap still increases but the width of the probability function
increases again which is due to the changing non-orthogonality of the phase
states for changing number of atoms, see Eq.~(\ref{eq:nonorth}).

\begin{figure}[tb]
\infig{20em}{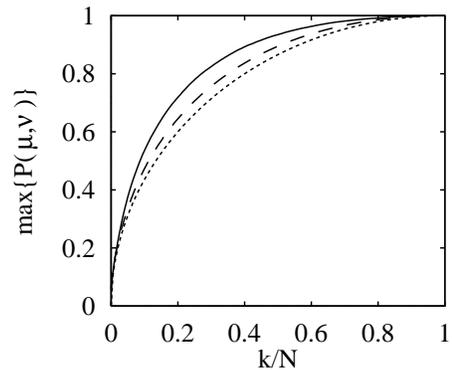}
\caption{Maximum of the probability function $P(\mu,\nu)$ after $k$
atom detections out of the $N=n_1+n_2$ initial atoms.
The solid curves correspond to the initial state 
$|\psi_0\rangle = |n_1=100,n_2=100\rangle$, the dashed curve to 
$|\psi_0\rangle = |n_1=200,n_2=50\rangle$, 
and the dotted curve is the maximum
of the approximate analytic solution, Eq.~(\protect\ref{eq:papprox}).
The first two cases are the results obtained from averaging over 2000
Monte-Carlo simulations.}
\label{fig:pphi2}
\end{figure}

In Fig.~\ref{fig:pphi2} we plot the {\em maximum\/} of the probability function
$P(\mu,\nu)$ as a function of the fraction of the detected atoms. The
approximate solution given by Eq.~(\protect\ref{eq:papprox}) is independent of
the total number $N=n_1+n_2$ of atoms initially in the two condensates and is a
quarter of a circle as a function of $k/N$. As already seen above, the
approach to an atomic coherent state starting from a pure number state is
in fact faster
than given by the approximation. Here we also plotted the numerical result for
an initial state with unequal atom numbers in the two condensates and we note
that also in this case the approximation by Eq.~(\protect\ref{eq:papprox}) is a
relatively good one.

Hence, we find that a significant fraction of the total number of atoms must be
detected in order that the system approaches an atomic coherent state. Thus the
time-scale for the preparation of such a state by detections is given by
$1/\gamma$ which widely differs from the time scale of $1/(N\gamma)$ 
found in the previous section to be relevant for the creation of interference
fringes.


\section{Generalisations of the model}
\label{sec:general}

\subsection{Imperfect detection}

We will now generalize our results from the previous sections to the case of
imperfect atom detection. Assuming that the detector efficiency is $\eta<1$ we
expect that after $k$ atoms have been lost from the condensates only $\eta k$
out of these are detected in the interferometric setup and thus contribute
to the build-up of coherence between the two condensates, whereas the remaining
$(1-\eta)k$ atoms are simple losses from either of the two condensates.

Hence, under the assumption that all detected atoms are found at the same
position (as in the previous sections) and the undetected atoms are coming
with the same probability from either of the two condensates, the state after
$k$ atoms have been lost from the condensates is approximately
\bea
|\psi_k\rangle &=& (a+b)^{\eta k} a^{\xi k} b^{\xi k} 
        |n,n\rangle \nonumber \\
     &=& \frac{n!}{(n-\xi k)!} (a+b)^{\eta k} |n-\xi k,n-\xi k\rangle
\eea
[where $\xi = (1-\eta)/2$] instead of the one given in Eq.~(\ref{eq:psikeq}).
Thus, in our earlier results, Eqs.~(\ref{eq:bapprox}) and (\ref{eq:papprox}),
we only have to substitute $n\rightarrow n-\xi k$ and $k\rightarrow \eta k$ to
obtain the approximate results for imperfect detector efficiency
\bea
\beta_k & \approx & e^{-1/(\eta k)}, \label{eq:bimp}\\
P(\phi) & \approx & \sqrt{1-\left(1-\frac{\eta k}{2n-k+\eta k}\right)^2}
   e^{-\phi^2 \frac{\eta k(2n-k)}{2n-k+\eta k}}.
\eea
As in the previous sections, comparison of these analytic approximations with
exact numerical solutions shows excellent agreement for the
visibility $\beta$ and an actually faster approach towards coherent states as
predicted by this approximation for $P(\phi)$.

Eq.~(\ref{eq:bimp}) shows that the only effect of imperfect detection on the
creation of interference fringes is that the number of atoms detected per unit
time is decreased and and thus it takes more time to detect the same number of
atoms as for perfect detectors. The effect of losses of atoms from individual
condensates does not seem to have any influence even if this changes the
relative atom number in the two condensates. However, since only a few atoms
need to be detected in order to build up the interference, the fluctuations
of the condensate occupation numbers remain very small compared to the
total number of
atoms and thus the maximum visibility is very close to the one for the
unperturbed initial state.

The situation is more complicated for the approach of the system state towards
an atomic coherent state (or a phase state) because, not only the number of
detected atoms, but also the total number of atoms left in the system plays an
important role, so that $P(\phi)$ depends on $k$ and $n$. The preparation of an
atomic coherent state also occurs on a larger time scale than for $\eta=1$
but the dependence on $\eta$ is not simple.


\subsection{Proper time evolution}

Another possible generalisation of our model is to investigate the system
evolution as a function of time instead of the number of detected atoms. Given a
constant decay rate of $\gamma$ for individual atoms from the condensates the
actual decay processes still occur in a probabilistic manner. Thus after a
certain amount of time the number of atoms remaining in the system is
uncertain. However, since the condensate decay follows an exponential law the
{\em mean\/} number of remaining atoms is
\be
\langle n\rangle(t) = N e^{-2\gamma t}
\ee
and so the mean number of detected atoms is
\be
\langle k\rangle(t) = N \left( 1-e^{-2\gamma t} \right).
\label{eq:meank}
\ee
We can generalise the results of the previous sections to incorporate this time
evolution by
simply replacing the number $k$ of detected atoms in 
Eqs.~(\ref{eq:bapprox}) and (\ref{eq:papprox}) by its mean value according
to Eq.~(\ref{eq:meank}). 

Using this assumption we compared 
the exact numerical results for the cases of the system
evolution versus number of detections and versus time.
We found that the
visibility is established slightly more slowly in the latter case. Let us consider,
for example, the state of the system after such a time $t$ that 
$\langle k\rangle(t) = 1$. At this time there is still a significant probability
that no atom was detected which greatly decreases the average visibility. On the other
hand there is a certain probability that two or three atoms were detected
which increases the average visibility, but since the difference between
$\beta_0$ and $\beta_1$ is much larger than the difference between $\beta_1$
and $\beta_2$ [see Eqs.~(\ref{eq:beta1}), (\ref{eq:beta2})] the former term
dominates and the average visibility is smaller than $\beta_1$. However, this
difference between the results for the two models decreases when more atoms are
detected and, starting from 100 atoms in each of the condensates, after $k=10$
detections the numerically found difference is already down to 1.7\%. A similar
agreement is also found for the maximum overlap of the system state with
atomic coherent states in the two models. We may thus conclude that our
approximate analytic solutions also describe the time evolution of the system
if one substitutes $k\rightarrow \langle k\rangle(t)$.


\subsection{Effect of collisions}

So far we have assumed the idealized case of noninteracting particles in the
condensates, so that our system was completely analogous to a system of photons in
two high-quality cavities from which the photons decay and interfere. All of our
previous results apply to this case as well \cite{Molmer}.

However, it is well known that atomic collisions play an important role in the
context of Bose-Einstein condensation \cite{Burt,Lewenstein,Javanainen2}. 
We will discuss the modifications to our
results in this case in the following. To this end, the free time evolution of
the system between two quantum jumps now has to be replaced by the time
evolution due to the collisional Hamiltonian which in our simple model of two
single-mode condensates is \cite{Wong,Sinatra,Steel}
\be
H = \kappa \left[(a^\dagger a)^2 +  (b^\dagger b)^2 \right].
\label{eq:ham_k}
\ee
The action of this Hamiltonian is to give different time dependent phases to the
various number states of the quantum state of the system of, for example, an
atomic coherent state. This dephasing gives rise to a time dependent ``decay'' of
the coherence and therefore of the conditional visibility $\beta$.
This decay of the coherence counteracts the creation of coherence due to atom
detections and thus prevents the system of reaching a state of maximum
visibility.

\begin{figure}[tb]
\infig{20em}{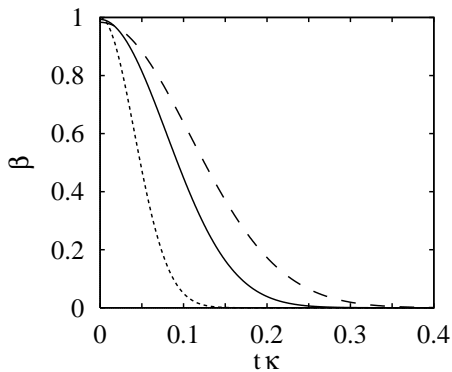}
\caption{Time evolution of the conditional visibility $\beta$ for different
quantum states including collisions of atoms in the condensates. The initial
quantum state is a state after 50 detections from the number state 
$|100,100\rangle$ (solid line), the state after 50 equal detections
$(a+b)^{50}|100,100\rangle$ (dashed), and a phase state 
$|\phi\rangle_{N=150}$ (dotted), respectively.}
\label{fig:beta_kt}
\end{figure}

First we study the evolution of various quantum states under the action of
the Hamiltonian (\ref{eq:ham_k}) without atomic decays. Let the system initially
be in an
atomic coherent state with equal mean atom number $N$ in the two condensates, 
$|\psi\rangle = |\mu,\mu\rangle_N$. The time evolution of the
visibility $\beta$, given by Eq.~(\ref{eq:beta}) can then be evaluated analytically as
\be
\beta(t) = \left[\cos(2\kappa t)\right]^{N-1} 
    \approx e^{-2\kappa^2 t^2 (N-1)}
\label{eq:betacoh}
\ee
where the latter approximation holds for small times $t \ll 1/\kappa$. Note that
the exact result of Eq.~(\ref{eq:betacoh}) predicts the well known
\cite{Castin,Sinatra,Steel,Wright} 
revivals of the visibility after times which are multiples of
$\pi/(2\kappa)$.

As noted, however, in the preceding sections,
the atomic coherent states
are in general not a good approximation to the state of the system 
until a significant number of atoms have been detected.
A better approximation is provided by considering the evolution of the initial
state $|\psi_k\rangle = (a+b)^k |n,n\rangle$. Using again the approximation of
Eq.~(\ref{eq:binapprox}) we obtain for this case
\be
\beta(t) \approx e^{-1/k} 
   \exp\left\{-2\kappa^2 t^2 \frac{k(2n-k-1)}{4n-k-2} \right\}.
\label{eq:beta_keq}
\ee
In the limit of $k \ll n$, we thus find that $\beta$ decays as
$\exp(-\kappa^2 t^2 k)$ which is much slower than the decay for an
atomic coherent
state (\ref{eq:betacoh}). For $k\rightarrow 2n$ the two expressions converge
since the state $|\psi_k\rangle$ approaches an atomic coherent state in this
case.

In Fig.~\ref{fig:beta_kt} we compare $\beta(t)$ for an atomic coherent state, the
state $|\psi_k\rangle$, and the numerical result for a state after 50 detections
from an initial number state $|100,100\rangle$ (all of these contain a
total number of 150 atoms). We see that in any case the collision-induced 
decay can be well
described by a Gaussian. The decay obtained for the state with simulated
detections is faster than the one of Eq.~(\ref{eq:beta_keq}) which agrees with
our finding of Sec.~\ref{sec:interference} that an atomic coherent state is
approached faster with arbitrary detections than if all detections occur at
the same position.

We will now use these results to derive an approximate analytic expression for
the visibility $\beta$ after $k$ detections including the effects of atomic
collisions. Let us assume that at a
given time $t_0$ exactly $k$ atoms have been detected from an initial number
state $|n,n\rangle$ and that the visibility is $\beta = \beta_0$. Then
the probability of detecting the next atom at time $t_0+t$ is given by
\be
P(t) = 2 n_0\gamma e^{-2n_0\gamma t}
\ee
where $n_0=2n-k$ is the number of atoms left in the system at time $t_0$. Thus,
if we write the decay of the visibility as 
\be
\beta(t_0+t) = \beta(t_0) e^{-t^2/\tau^2},
\ee
where $\tau$ is given in Eq.~(\ref{eq:beta_keq}), then the visibility at the time
immediately {\em before\/} the next atom detection is on average given by
\bea
\langle \beta(t_0+t) \rangle &=& \int_0^{\infty}dt\,\beta(t_0+t)P(t) 
   \nonumber\\
&=& \beta(t_0) 2n_0\gamma\tau 
    \frac{\sqrt{\pi}}{2} e^{n_0^2\gamma^2 \tau^2}
    \left[1-\Phi(n_0\gamma\tau ) \right],
\eea
where $\Phi$ denotes the error function. Let us now assume that the 
system is in steady state between the creation and the decay of $\beta$, 
that is, the following detection increases the visibility again to its
value $\beta(t_0)$ at time $t_0$. Hence, writing
\be
\beta(t_0) = e^{-1/k_0} \approx 1 - \frac{1}{k_0}
\ee
and using our previous result for the increase of $\beta$ with the number of
detections, Eq.~(\ref{eq:bapprox}), we obtain the following condition
\be
\left( 1 - \frac{1}{k_0} \right) \langle \beta(t_0+t) \rangle 
   = 1 - \frac{1}{k_0-1}
\ee
with the solution
\be
k_0 = 1 + \frac{1}{\sqrt{1-\langle \beta(t_0+t) \rangle}}
\ee
and thus the steady state visibility is
\be
\beta_{\text{st}} = \frac{1}{1 + \sqrt{1-\langle \beta(t_0+t) \rangle}}.
\label{eq:beta_stat}
\ee

\begin{figure}[tb]
\infig{20em}{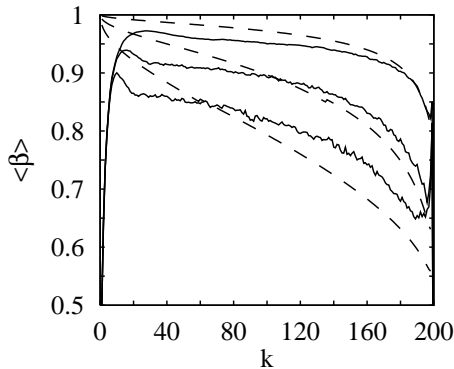}
\caption{Visibility $\beta$ vs number of detected atoms $k$ from an initial
number state $|100,100\rangle$ including atom collisions. Solid lines
are exact results (averages of 1000 Monte-Carlo simulations), dashed lines are
the corresponding analytic approximations given by 
Eq.~(\protect\ref{eq:beta_stat}). The
parameters are (from top to bottom) $\kappa = 0.5\,\gamma$,
$\kappa = 2\,\gamma$, and $\kappa = 5\,\gamma$.
}
\label{fig:beta_kk}
\end{figure}

We compare this result with the results of Monte-Carlo simulations in
Fig.~\ref{fig:beta_kk} for different values of $\kappa$. From this we note that
the approximation yields values of $\beta$ that are too large. This is
especially true for small
numbers $k$ of detected atoms ($k<n$). This was to be expected since (i) the
approximation was based on the assumption of a steady state whereas it takes a
certain number of detections to reach this state in the simulations and (ii) in
the discussion of Fig.~\ref{fig:beta_kt} we have already seen that the
actual decay of the
visibility occurs faster than predicted by Eq.~(\ref{eq:beta_keq}) which in
turn
decreases the steady state value. On the other hand, the approximation yields
values of $\beta$ below the numerical results for values of 
$\beta < 0.8$ where
we know that the increase of the visibility due to detections is faster than
given by Eq.~(\ref{eq:bapprox}).

\begin{figure}[tb]
\infig{20em}{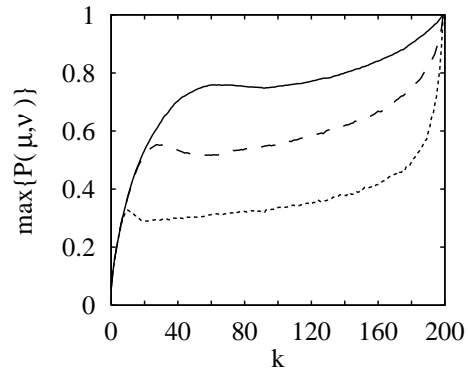}
\caption{Maximum overlap $\mbox{max}\{P(\mu,\nu)\}$ 
vs number of detected atoms $k$ from an initial
number state $|n_1=n_2=100\rangle$ including atom collisions with collision
rate $\kappa = 0.1\,\gamma$ (solid line),
$\kappa = 0.5\,\gamma$ (dashed), and $\kappa = 5\,\gamma$ (dotted).
Each curve is obtained from averaging over 1000 Monte-Carlo simulations.
}
\label{fig:beta_kp}
\end{figure}

Finally we will investigate the effects of atomic collisions in the condensates
on the creation of atomic coherent states. To this end we calculate the
maximum overlap of the time evolved state
\be
|\psi_k(t)\rangle = e^{-iHt} |\psi_k\rangle,
\ee
where $|\psi_k\rangle$ is the state after $k$ equal detections from the number
state $|n,n\rangle$ as used before, with the atomic coherent states.
Following the same steps as in Sec.~\ref{sec:state} one obtains
\be
\mbox{max}\{P(\phi,t)\} = \frac{\mbox{max}\{P(\phi,t=0)\}}{\sqrt{
      1+\left[ t\kappa k(2n-k)/(2n) \right]^2
   }}
\ee
where $\mbox{max}\{P(\phi,t=0)\}$ is the maximum overlap at time $t=0$ given by
Eq.~(\ref{eq:papprox}).
Hence, for $k\ll n$ this maximum overlap decays on a time scale of 
$t\sim 1/(\kappa k)$ which is much faster than the time
scale of the decay of the visibility where $t\sim 1/(\kappa \sqrt{k})$,
Eq.~(\ref{eq:beta_keq}). This, together with the much longer time scale necessary
to create a significant overlap with coherent states, shows that atomic
collisions have a much larger effect on $P(\phi,t)$ than on the visibility.
The time scales for the decay and the build-up of this state overlap also
prevent the system from reaching a steady state and thus no analytic
approximation analogous to Eq.~(\ref{eq:beta_stat}) can be found. Hence we have
to rely on numerical simulations in this case, see Fig.~\ref{fig:beta_kp}.

We note that for small numbers of detected atoms the maximum overlap increases
close to the case of $\kappa=0$. For larger $k$, corresponding to a broader
distribution of the relative atom number between the two condensates, the
dephasing due to the atom collisions starts to dominate and significantly
reduces $\mbox{max}\{P(\mu,\nu)\}$ even for small values of $\kappa$.
However, when most of the atoms have been detected ($k\rightarrow 2n$) the
function approaches unity as all one-particle states and the final vacuum
state are exact atomic coherent states.


\section{Conclusions}

In this work we have studied in detail the creation of coherence between two
initially uncorrelated
Bose-Einstein condensates according to the detections of individual atoms in an
interferometric setup. Our main finding is that first order coherence, as
observed in the interference fringes of the condensates, is established on a time
scale corresponding to the detection of only a few atoms, that is, within times
$t\sim 1/(N\gamma)$, where $N$ is the total number of atoms in the condensates
and $\gamma$ is the single-atom decay rate. On the other hand, high-order
coherence,
which in this article we describe by the overlap of the quantum state of the 
system with coherent states, is created by detecting a certain (large) {\em
fraction\/} of the condensed atoms, so the time scale for this develop 
is given by 
$t\sim 1/\gamma$. These results were obtained from approximate analytic
solutions as well as from exact numerical simulations. 

We have also investigated the effect of atomic collisions on these results. 
In this case the dephasing of the quantum state of the system due to the
collisions counteracts the entangling effect of the atom detections. However,
according to the widely different time scales a good visibility of the
interference fringes can be maintained whereas the atomic collisions
effectively prevent the preparation of a coherent state even for relatively
small collision rates.


\acknowledgments

This work was supported by the United Kingdom Engineering and Physical Sciences
Research Council.


\begin{appendix}

\section{Atomic coherent states}
\label{appendix}

The atomic coherent states \cite{states} were defined in Eq.~(\ref{eq:qcs})
to be
\be
|\mu,\nu\rangle_N = \frac{1}{\sqrt{N!}} 
   \left( a^\dagger \mu + b^\dagger \nu\right)^N |0,0\rangle,
\ee
where $|\mu|^2 + |\nu|^2 = 1$. These are states with precisely $N$ atoms shared
between the two condensates and can be expressed as an entangled superposition
of all product number states for which the total number is $N$:
\be
|\mu,\nu\rangle_N = \nu^N \sum_{n=0}^N \bin{N}{n}^{1/2}
   \left(\frac{\mu}{\nu}\right)^n |n,N-n\rangle.
\ee
The number of atoms in one of the condensates, therefore, has a binomial
distribution.
Different atomic coherent states to the same number of atoms $N$ are in general
not orthogonal to each other, but have a nonvanishing scalar product
\be
{}_N\langle \mu',\nu'|\mu,\nu\rangle_N = \left( 
      \mu\mu'^* + \nu\nu'^*
   \right)^N.
\label{eq:nonorth}
\ee

The atomic coherent states are related to the familiar two-mode coherent
states $|\alpha,\alpha'\rangle$ by
\be
|\alpha,\alpha'\rangle = e^{-(|\alpha|^2+|\alpha'|^2)/2}
   \sum_{N=0}^{\infty} \sqrt{\frac{(|\alpha|^2+|\alpha'|^2)^N}{N!}}
   |\mu,\nu\rangle_N
\ee
with
\bea
\mu &=& \frac{\alpha}{\sqrt{|\alpha|^2+|\alpha'|^2}}, \\
\nu &=& \frac{\alpha'}{\sqrt{|\alpha|^2+|\alpha'|^2}}.
\eea
Clearly, the restriction of a two-mode coherent state to the $N$ particle
subset of states is an atomic coherent state. 

The atomic coherent states satisfy the following useful identities
\bea
a |\mu,\nu\rangle_N & = & \sqrt{N} \mu |\mu,\nu\rangle_{N-1}, \\
b |\mu,\nu\rangle_N & = & \sqrt{N} \nu |\mu,\nu\rangle_{N-1}.
\eea
From these it immediately follows that the expectation values of $a$ and $b$ in
an atomic coherent state are zero so that neither condensate exhibits a
prefered phase. It is also clear that the mean number of atoms in the
condensates $a$ and $b$ are $N|\mu|^2$ and $N|\nu|^2$, respectively, and that
the expectation value of $a^\dagger b$ is $N\mu^*\nu$. It follows that the
conditional visibility for an atomic coherent state is
\be
\beta_c = 2|\mu||\nu|
\ee
which assumes its maximum value of unity if and only if the two condensates
have the same mean occupation number.
Restricting considerations to equal mean occupation numbers leads to the phase
states \cite{Castin,Sinatra} defined in Eq.~(\ref{eq:phs}). Further properties
of the atomic coherent states can be found in the literature \cite{states}.
\end{appendix}


\end{document}